\documentclass[conference]{IEEEtran}
\IEEEoverridecommandlockouts
\usepackage{cite}
\usepackage{amsmath,amssymb,amsfonts}
\usepackage{graphicx}
\usepackage{textcomp}
\usepackage{xcolor}
\def\BibTeX{{\rm B\kern-.05em{\sc i\kern-.025em b}\kern-.08em
    T\kern-.1667em\lower.7ex\hbox{E}\kern-.125emX}}

\usepackage{booktabs}
\usepackage{enumitem}
\usepackage{pgfplots}
\pgfplotsset{compat=1.18}

\usepackage{tabularx}
\newcolumntype{L}[1]{>{\raggedright\arraybackslash}m{#1}}
\newcolumntype{C}[1]{>{\centering\arraybackslash}m{#1}}
\newcolumntype{R}[1]{>{\raggedleft\arraybackslash}m{#1}}
\newcolumntype{Y}{>{\centering\arraybackslash}X}
\usepackage{multirow}
\usepackage{makecell}

\usepackage{subcaption}

\begin{document}

\title{AdaptoNet: Modular Foundation--Adaptive Neural Networks for Cyber-Physical Attack Detection in Power Grids}

\author{%
\IEEEauthorblockN{Anissa Elias\IEEEauthorrefmark{1}, Jennifer Rogers\IEEEauthorrefmark{2}, Hui Lin\IEEEauthorrefmark{1}, and Yan (Lindsay) Sun\IEEEauthorrefmark{1}}
\IEEEauthorblockA{%
\IEEEauthorrefmark{1}Dept. of Electrical and Computer Engineering, University of Rhode Island, Kingston, RI, USA\\
\IEEEauthorrefmark{2}Dept. of Electrical Engineering and Computing, United States Coast Guard Academy, New London, CT, USA\\
\{anissa\_elias, huilin, yansun\}@uri.edu, \quad Jennifer.A.Rogers@uscga.edu}%
}

\maketitle

\begin{abstract}
Cyber attacks on the power grid combine physical disruptions with compromised data to destabilize cyber-physical systems. We demonstrate that data denial attacks, where adversaries block measurements in a targeted region while triggering a line outage, reduce detection performance by more than 86\%, rendering standard data-driven methods ineffective. We propose AdaptoNet, a modular neural network that adapts to measurement availability through conditional controls. AdaptoNet pairs a frozen foundational module trained on complete data with a trainable adaptive module, conditioned on a binary measurement-availability vector, enabling the model to distinguish between denied and anomalous data without retraining the foundational module. Evaluated across four IEEE test systems (30-, 39-, 57-, and 118-bus) under in-region attacks blocking up to 20\% of measurements, AdaptoNet recovers F1 from below 12\% to above 81\%, an approximate sevenfold improvement approaching the 89\%--99\% baseline with complete measurements.
\end{abstract}

\vspace{-3pt}
\vspace{-3pt}
\section{Introduction}

Off-the-shelf computing and network technologies elevate power grids into a highly coordinated cyber-physical system (CPS). Leveraging such interdependency between physical processes and computing devices, malicious actors can launch complicated \textit{cyber-physical attacks}, amplified by the ability to coordinate physical disruptions (e.g., line outage) and
measurement data compromise~\cite{zhang2021smart}.  
The 2015 attacks on Ukrainian power plants introduced coordinated cyber intrusions causing outages affecting 225,000 residents, followed by voluminous phone calls preventing energy companies from accessing customers' reporting outages~\cite{lee2016analysis}. The 2022 Sandworm campaign further illustrated the evolving threat, combining missile strikes against physical infrastructure with concurrent cyber operations to mask runtime operational status~\cite {proska2023sandworm}. 

These coordinated cyber-physical attacks can significantly degrade the effectiveness of existing security methods, which heavily rely on measurement data collected through IP-based industrial control networks. Model-based methods use optimization problems,
e.g., weighted-least-square (WLS), over power grids’ physical models. They generally require sufficient measurements to ensure observability and incur high computational latency. To address these shortcomings, data-driven methods that heavily rely on machine learning models are increasingly promising because of their ease of deployment, low runtime overhead, and limited reliance on grid physical models. However, the compromised data present a misaligned distribution from the training data, leading to a significant performance downgrade of the machine learning models (see Fig.~\ref{fig:attack_impact} for details). Making things worse, imputation methods, whether discriminative~\cite{stekhoven2011missforest} or generative~\cite{yoon2018gain}, assume randomness in missing data, which may not hold in cyber-physical attacks, where missing measurements are closely coordinated with physical disruptions.

To address this gap, we propose AdaptoNet, a modular neural network architecture that adapts to measurement availability. The originality in the architecture consists of two modular components: a foundation module and an adaptive module. \textbf{The foundation module} is trained on fully observed data to approximate the physical model that governs the runtime behaviors of a target grid, enabling anomaly detection with complete measurements. Once trained, it is frozen and combined with \textbf{the adaptive module}, trained separately to learn the conditional mapping from partially observed measurements (indicated via mask variables) to anomaly likelihoods. 

This modular design is structurally related to approaches showing great success in AI applications. Its structure is inspired by diffusion models, which have ``frozen'' and ``trainable'' components with mirrored internal infrastructure~\cite{zhang2023adding}. AdaptoNet differs in that the foundational and adaptive modules can share very different infrastructures, allowing flexible integration of pre-trained or domain-specific modules and enabling a plug-and-play approach. Inspired by fine-tuning in language models, AdaptoNet differs by explicitly defining the parameter space in the adaptive module to address missing measurement scenarios.

Specifically, our contributions are:
\begin{itemize}[leftmargin=*,topsep=2pt, itemsep=1pt, parsep=0pt]
\item A modular neural network architecture that separates general grid knowledge (foundational module) from attack-specific adaptation (adaptive module), preserving learned grid knowledge when adapting to new threat models.

\item A conditioning mechanism that encodes measurement availability as a spatial indicator, providing the model with knowledge of what buses the adversary has compromised.
\item Evaluation across four IEEE test systems (30-, 39-, 57-, and 118-bus) showing that this approach recovers detection F1 from below 12\% to above 81\% under attacks blocking up to 20\% of measurements.
\end{itemize}


\section{Related Works}\label{literature}

\textbf{Missing Data Imputation.}
Imputing missing data has been an active research topic for machine learning models. Discriminative methods leverage models such as random forest to estimate missing values from observed data distributions~\cite{stekhoven2011missforest}, while generative methods directly produce missing values conditioned on observed data, including variational autoencoders~\cite{collier2021vae} and generative adversarial networks~\cite{yoon2018gain}. In power systems, these approaches inform pseudo-measurement crafting algorithms~\cite{xu2025pseudo}. These imputation methods assume randomness in missing data, e.g., missing completely at random (MCAR) or missing at random (MAR); they recover missing data according to probability distribution of existing data. Those assumptions may not hold under cyber-physical attacks. More importantly, AdaptoNet's objective is not to recover missing data. Instead, it uses a new modular structure to make machine learning models resilient against missing data in cyber-physical attacks.

\textbf{Data-Driven Line Outage Detection.}
In existing data-driven methods for line outage detection and localization~\cite{Learning_to_infer, Lineoutages_PMU, Hidden_markov, sysID, previous_work}, convolutional neural networks (CNNs) have proven particularly effective~\cite{CNN_LSTM, CNN_faultdiagnosis, Zhuo_CNN_2024}. However, these studies assume models are trained on data that represents the testing conditions; they do not generalize well to unknown operating conditions, as might be experienced during a cyber-physical attack. To address this, transfer learning has been proposed to enhance generalization of static models to evolving data~\cite{shakiba2022transferlearningfaultdiagnosis}, and prediction-based data augmentation has demonstrated promise in improving line outage classification~\cite{rogers2024prediction}. However, these methods rely on accurate measurements at test time, which may not be available during cyber-physical attacks.

\textbf{Cyber-Physical Attack Models and Detection.}
Studies in~\cite{zhang2021smart, cyber_attack_types} discuss attacks involving physical disruptions combined with data manipulation, which can vary in complexity based on attackers' level of network access. 
In~\cite{coordinated_attack}, Lai et al. use an optimization method to identify critical data, whose protection can mitigate the impact of missing other data on cyber-physical attack detections. Studies in~\cite{After_attack_Bayes, lineoutage_PMU_joint_attack, soltan2017react, ren2025faultlocalizationstateestimation} share a similar attack scenario considered in this work, measurements blocked in the same regions where physical disruptions occur. However, these solutions basically transfer the cyber-physical attack detection into missing data imputation problems. For example, studies in~\cite{After_attack_Bayes, ren2025faultlocalizationstateestimation} use physics-based models of power grids to estimate missing data in attacked zones from the data outside the zone while \cite{lineoutage_PMU_joint_attack} uses linear minimum mean square estimation for the same objective. Studies in~\cite{soltan2017react} extends~\cite{After_attack_Bayes} to approximate attacked regions and then estimate missing data. The performance of those detection methods heavily rely on the accuracy of the recovered missing data. Meanwhile, optimization-based methods can generally suffer the curse of dimensionality. Instead, AdaptoNet provides an end-to-end solution to detect cyber-physical attacks, aiming to improving the resilience of data-driven methods. 

\textbf{Conditioning in Adjacent Domains.}
To the best of our knowledge, AdaptoNet is the first work to use conditional control for cyber-physical attack detection in smart grids, inspired by data conditioning used to improve model performance in other domains. In~\cite{Conditioning_image_generation}, the authors enhance image composition by conditioning generation on user-provided text-prompts, offering greater spatial control. In~\cite{mcroberts2018improving}, conditioning improves hurricane-related power outage predictions by identifying key explanatory variables through a generalized model and refining predictions in a two-step procedure. In~\cite{nn_conditioning}, conditioning integrates load forecasting and demand charge threshold optimization end to end, mitigating sensitivity to forecast errors. Collectively, these works demonstrate how augmenting input data with relevant features can boost baseline model accuracy, the same principle we adapt to the CPS attacks. 

\section{Threat Model}\label{subsec:threat}

Power grid CPS security faces a unique challenge: the tight coupling between cyber and physical layers means that compromise of either layer can cascade into the other. Following taxonomy in~\cite{lin2016safetycritical}, we can classify threat models considered in existing work by entry points. In the first category, attacks exclusively target measurements (often called false or bad data-injection attacks) to mislead control algorithms by corrupting state estimates~\cite{liu2011fdia}. In the second, attacks exclusively manipulate control commands to directly compromise runtime system states~\cite{lin2018runtime}. 

We consider \textit{cyber-physical attacks} that advance these conventional models by strategically coordinating measurement compromise with physical disruption. These attacks exploit the fundamental dependency of grid monitoring on measurement integrity; both conventional monitoring applications (e.g., state estimation, contingency analysis) and newly developed ML-based security solutions that require consistent and sufficient measurements to maintain acceptable performance may become ineffective against these attacks. 

We make two assumptions that ground this concept in feasible attack capabilities. (i) \textbf{Measurement blocking}: adversaries block measurements rather than craft specific values, since conforming to power system physics is computationally prohibitive~\cite{kosut2011malicious}; blocking is achievable via link-flooding attacks that congest critical communication paths~\cite{rezazad2019detecting}. (ii) \textbf{Line outage}: the physical disruption is a line outage, achievable remotely on intelligent electronic devices, rather than an electrical fault mitigated by hardwired safety systems. Line outages introduce stealthy perturbations propagating through the grid, requiring wide-area monitoring for detection.

Under these assumptions, the spatial relationship between the denied region and the outage location defines distinct attack strategies. In an \textit{in-region attack}, the line outage occurs within the data-denied region. In an \textit{out-region attack}, the outage occurs outside the denied region, decoupling spatial correlation. A \textit{random attack} generalizes both by selecting the outage location independent of the denied region. In this paper, we specifically focus on \textbf{in-region attacks}. This strategy represents a low-capability but high-impact threat, which is also consistent with~\cite{After_attack_Bayes, lineoutage_PMU_joint_attack, soltan2017react}. With regional access to block data and trip a single line, attackers can easily cause data denial that can reduce detection performance by over 86\% (see Figure~\ref{fig:attack_impact}), as this area-based suppression removes the most informative measurements, data from buses closest to the disruption. However, the spatial correlation between the denied region and the outage location can narrow investigative efforts, striking a tradeoff that attackers accept for operational simplicity. It is worth noting that AdaptoNet's design is not restricted to these specific attack strategies, but can be extended to more complicated scenarios, which we leave as future work (see Section~\ref{subsec:discuss} for details).



\section{AdaptoNet Design}\label{conditioning_soln}

In this section, we first present a motivation example showing how cyber-physical attacks downgrade state-of-the-art data-driven anomaly detections. Then, we present the infrastructure and training procedures in AdaptoNet, inspired by strategies that demonstrate great success in general AI applications.

\subsection{Motivation}

As a motivation example, we implement the in-region line outage attack on four IEEE test systems. For each system, we randomly select a region including the closest 10\% of buses and block all measurements from those buses and connecting transmission lines. We then perform a single-line outage within the region, simulated in MATPOWER~\cite{matpower}.

We adopt the CNN-based line outage detection method from~\cite{rogers2024prediction}, which processes voltage phasor measurements (magnitudes and angles at each substation) and outputs a multi-label vector indicating the status of each transmission line. To enable testing under data denial, blocked measurements are replaced with zeros. This strawman approach preserves input dimensions but confuses the neural network, which cannot distinguish zero-valued measurements from genuinely absent data.

\begin{figure}[htbp]
    \centering
    \includegraphics[width=0.85\columnwidth]{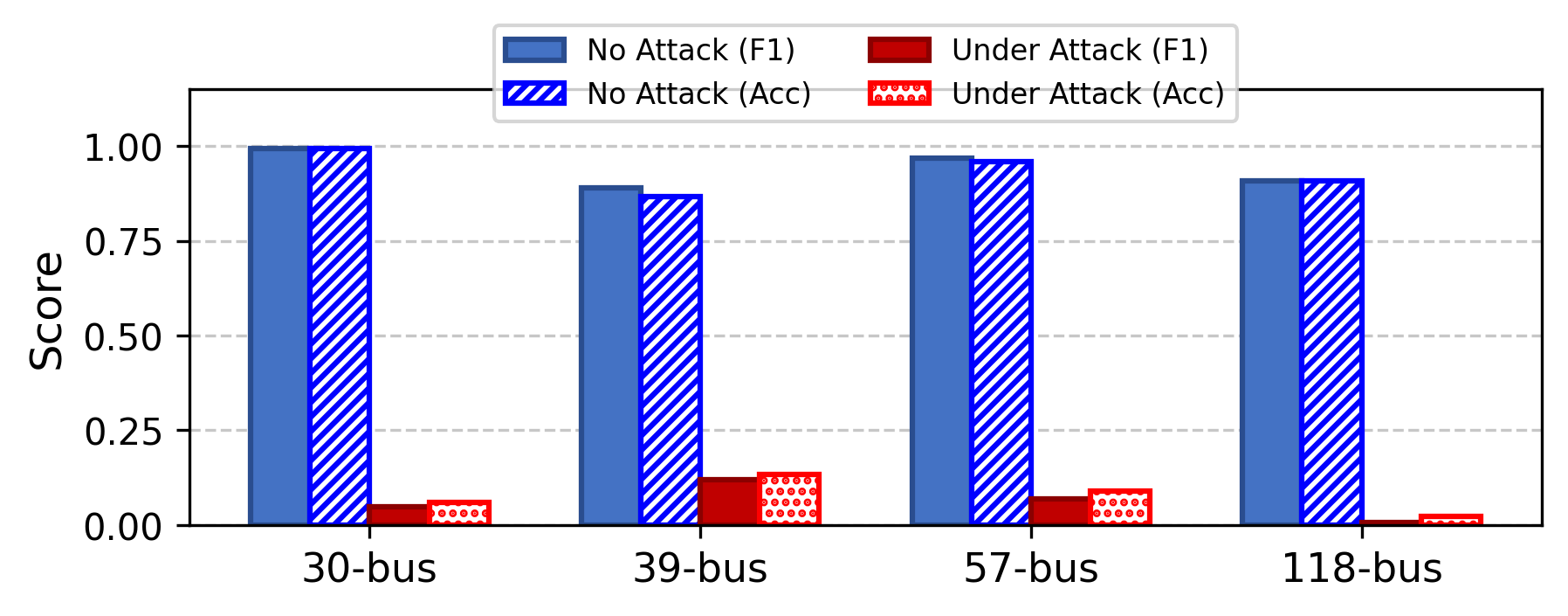}
    \vspace{-5pt}
    \caption{Impact of in-region cyber-physical attacks (10\% buses compromised) on line outage detection. F1 collapses from 89--99\% to below 12\% and accuracy collapses from 87--99\% to below 13\% across four IEEE test systems.}
    \label{fig:attack_impact}
    \vspace{-5pt}
\end{figure}

As shown in Fig.~\ref{fig:attack_impact}, F1 scores collapse from 89\%--99\% to below 12\% across all four systems, demonstrating that even simple data denial renders data-driven detection ineffective. The 118-bus system is most severely affected (0.6\% F1), as its larger topology distributes outage signatures more widely. This severe degradation motivates the need for detection methods that can adapt to measurement availability. 

\subsection{Architecture}

\begin{figure}[htbp]
 \centering
 \includegraphics[width=2.6in]{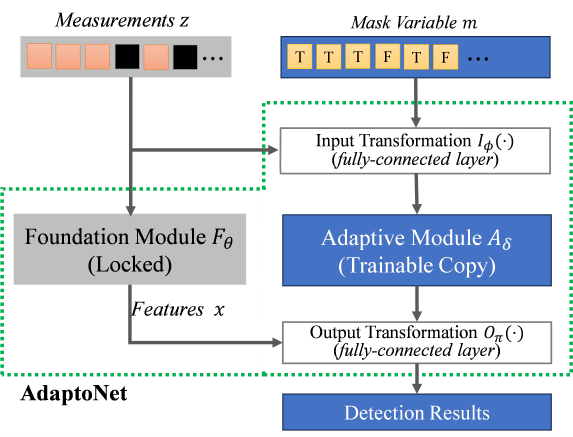}
 \caption{AdaptoNet architecture with foundational and adaptive modules.}
 \label{conditioning_model}
\end{figure}

AdaptoNet uses a modular neural network infrastructure, with a foundational module and a trainable adaptive module (Fig.~\ref{conditioning_model}), inspired by conditional control mechanisms in diffusion models where locked and trainable network components are combined with conditional inputs~\cite{zhang2023adding}. In diffusion models, the locked copy captures general features of images, e.g., shapes, edges, and object types, while the trainable copy correlates external conditional variables to the details e.g., hair style and eye colors. In AdaptoNet, the \textit{foundational module} $\mathcal{F}_{\theta}(\cdot)$ maps measurements $z$ to features $x = \mathcal{F}_{\theta}(z)$, capturing general knowledge, e.g., physical models, that governs various operating conditions of a power grid. Its parameters are frozen after initial training.

The \textit{adaptive module} integrates measurement availability (mask variable $m$) with measurements $z$ to produce detections resilient against data denial. Input and output transformations ($\mathcal{I}_{\phi}$ and $O_{\pi}$), implemented as fully-connected layers, bridge the foundational and adaptive modules:
\begin{equation}
x=\mathcal{F}_{\theta}(z) \label{eq:adaptonet1}
\end{equation}
\begin{equation}
y=O_{\pi}(A_{\delta}(\mathcal{I}_{\phi}(z, m)), x) \label{eq:adaptonet2}
\end{equation}
where $A_{\delta}$ is the adaptive module with parameters $\delta$, and $O_{\pi}$ fuses the adaptive and foundational outputs into the final detection, $y$, a vector indicating line status.

The modular split serves a specific function not just an architectural convenience. Simply appending a mask to the input of a single network would require the model to simultaneously learn grid physics and attack-specific adjustments, risking interference between these objectives. By freezing the foundational module, we guarantee that learned physical relationships are preserved regardless of how the adaptive module is trained. The foundational module provides a stable reference prediction from complete-data knowledge, while the adaptive module learns \textit{corrections} conditioned on what is missing. The output fusion layer then combines both signals.

This modular design supports two usage modes:
\begin{itemize}
\item \textbf{Mirroring Foundation and Adaptive Modules.} Both modules share the same architecture: the foundational module is trained on complete data, frozen, and its weights initialize the adaptive module, which is trained exclusively on attacked data with the conditioning mask.
\item \textbf{Using Third-party Foundation Module.} The foundational module can be a third-party pre-trained model (e.g., a deep-learning-based state estimator), with the parameterized input/output transformations bridging differences in data format. This flexibility enables integration of domain-specific or proprietary models as foundations for attack-adapted detection. 
\end{itemize} 

In this paper, we focus on the first option and mainly demonstrate the effectiveness of the modular framework. We can easily implement the second design option by changing input/output transformations in AdaptoNet. 
Following~\cite{After_attack_Bayes, lineoutage_PMU_joint_attack}, we also assume that we know the data denial region and the corresponding blocked data, which determines the mask variable. In practice, operators can determine the availability mask directly from the communication layer (e.g., missing Supervisory Control and Data Acquisition (SCADA) or Phasor Measurement Unit (PMU) reports). However, AdaptoNet's design allows easy extension to complex scenarios, including partial corruption, spoofed data, and data replay~\cite{soltan2017react}, when working collaboratively with existing security solutions. Instead of using binary values to indicate data availability, we can extend mask variables with floating points to include richer knowledge, e.g., the accuracy of the data values estimated by missing data imputation methods or their trustworthiness determined by intrusion detection systems. We leave it as future work.  

\subsection{Training}
\begin{figure}[!t]
 \centering
 \includegraphics[width=1.7in]{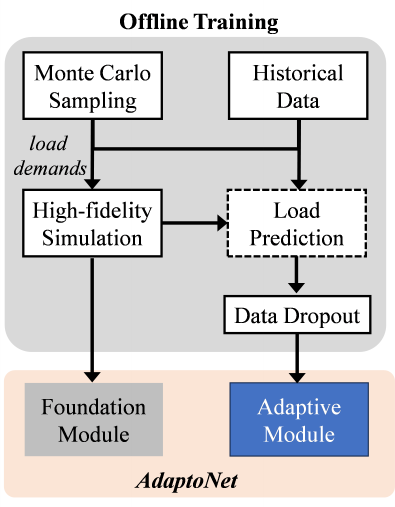}
 \caption{Two-stage training procedure for AdaptoNet.}
 \label{fig:training}
\end{figure}

We train AdaptoNet in a two-stage procedure (Fig.~\ref{fig:training}). This is inspired by foundational models used in generative AI applications, e.g., image creation or large language models (LLMs).  

\textbf{Stage 1: Foundational module training.} The foundational module is trained on historical data following the standard procedure for data-driven anomaly detection. The training dataset includes simulations of both normal operation and single-line outages for all transmission lines, using voltage phasor measurements generated via AC optimal power flow (AC-OPF). The foundational module converges to capture the general physical relationships governing grid behavior under complete measurement conditions.

\textbf{Stage 2: Adaptive module training.} After freezing the foundational module parameters, we switch to cyber-physical attack scenarios. Since the foundational module already captures realistic operating conditions, we rely on Monte Carlo sampling and attack injection in simulation to generate training data with controlled complexity. We intentionally align training complexity with module activation: the foundational module handles general physics while the adaptive module addresses increasingly complex attack patterns. 

Unlike LLMs, which leave a large footprint in parameter spaces and can struggle to determine parameters to optimize during fine-tuning, AdaptoNet's modular design explicitly separates these objectives. This introduces two benefits: (i) the foundational module remains frozen, preserving learned physics, and (ii) a training curriculum can be clearly defined by increasing the attack complexity.

\textbf{Training cost.} AdaptoNet includes approximately 373 thousand to 10 million parameters (30- to 118-bus) with both stages run on a single NVIDIA TITAN RTX GPU. Because of mechanical inertia, operational states of power grids change slowly. The foundational module trains in roughly 8--17 minutes per system (30- to 118-bus), and each adaptive module trains in a comparable time per region for 100 hours of operational data. Once trained, the foundational module is frozen, and fine-tuning of both modules can occur infrequently. 

\section{Evaluation}\label{conditioning_results}

\subsection{Experiment Setup}

Our evaluation uses realistic operational load profiles from the publicly available \textit{ACTIVSg2000} dataset~\cite{b20}, which contains hourly load time-series data spanning a full calendar year across 2,000 buses, derived from actual U.S. grid demand patterns. We select six non-overlapping 100-hour windows from this dataset to create six evaluation cases, each representing distinct seasonal and diurnal operating conditions. The load profiles are normalized to the bus capacities of each IEEE test system (30-, 39-, 57-, and 118-bus), preserving the realistic temporal correlations and demand variability present in the original data.

For each hour, we compute bus voltage magnitude and phasor angles by solving AC optimal power flow (AC-OPF) in MATPOWER~\cite{matpower}. Line outages are injected by removing individual lines and re-solving the AC-OPF, producing post-outage steady-state measurements as a wide-area monitoring system would observe. This pipeline ensures measurements reflect the nonlinear power flow physics of real grid behavior.

\subsubsection{Data Generation} For each of the six evaluation cases, we split 100 hours of load data into 80 hours for training the foundational module and 20 hours for testing. Training data includes simulation of normal operation and single-line outages across the 80 training hours, providing an even representation of normal and abnormal behavior. 

For the adaptive module, we create training data based on the proposed threat model. Similar to the foundational training data, the adaptive dataset includes normal system behavior, but only simulates line outages within the attacked region. For each region, the dataset contains normal scenarios and abnormal scenarios equivalent to the number of lines within the region. Attacked-region measurements are zeroed, and the availability mask is set accordingly. In summary, we generate six cases, each with 100 hours of operating data (80 training, 20 testing), with 3--20 attackable lines per region, depending on region type and system size.

We evaluate on four IEEE systems of increasing size: 30-bus (41 lines), 39-bus (46 lines), 57-bus (80 lines), and 118-bus (186 lines). The input consists of time-dependent tensors of bus voltage phasor measurements; the multi-label output vector indicates the operational status of each transmission line. Given the class imbalance inherent in single-line outage detection, we report recall, precision, and F1-score.

\subsubsection{Regional Selection} Based on our threat model, we create a graphical representation of each IEEE test system, where buses are nodes and transmission lines are edges. We traverse each node and create two types of regions using Dijkstra's algorithm to compute node-to-node distances. Type~\textit{A} regions include the selected bus and the closest 10\% of buses, creating a region of approximately four affected buses. Type~\textit{B} regions include the closest 20\%, creating a region of approximately seven affected buses. 

For each defined region, the simulated cyber-physical attack targets buses and lines within the region. The attack compromises data integrity, rendering bus measurements unusable, and induces a physical failure of a single transmission line within the attacked region. The attack scenario assumes limited attacker capability confined to the region.

\subsubsection{Non-Convergent Branch Handling} During data generation via AC power flow simulation, certain transmission lines produce non-convergent solutions when removed from the network. These usually occur when a line outage puts the target system into islands (e.g., loss of critical generators or tie-lines) that would likely trigger protective relay action before reaching steady state. These physically extreme contingencies can be detected directly based on topology information, making them less relevant to the data-driven detection scenario we study. Table~\ref{tab:excluded} lists the non-convergent branches per system. We employ a \textit{class mask} that excludes these lines from both loss computation and metric evaluation. 

\begin{table}[htbp]
\centering
\caption{Non-convergent branches excluded from evaluation.}
\label{tab:excluded}
\begin{tabular}{lcc}
\toprule
\textbf{System} & \textbf{Total Lines} & \textbf{Non-Convergent} \\
\midrule
IEEE 30-bus  & 41  & 7 (17\%) \\
IEEE 39-bus  & 46  & 11 (24\%) \\
IEEE 57-bus  & 80  & 25 (31\%) \\
IEEE 118-bus & 186 & 9 (5\%) \\
\bottomrule
\end{tabular}
\end{table}

\subsection{Evaluation Results}

\subsubsection{AdaptoNet Performance}
Table~\ref{tab:adaptonet_results} summarizes performance under both Type~A (10\% compromised) and Type~B (20\% compromised) regions. For each region, we select the decision threshold maximizing F1 on a held-out validation split (distinct from the test set) and report test metrics averaged across all testable regions and six evaluation cases. Per-region thresholds are used because the class balance varies across regions; the values reported here therefore isolate the detector's discriminative capacity from operational threshold selection. In deployment, where the attacked region is unknown, a single conservative threshold would be applied across all regions, so these per-region results represent an upper bound relative to a single-threshold deployment. 

\begin{figure*}[!htbp]
  \centering
  \begin{minipage}{0.48\textwidth}
    \centering
    \includegraphics[width=\textwidth, height=0.35\textwidth]{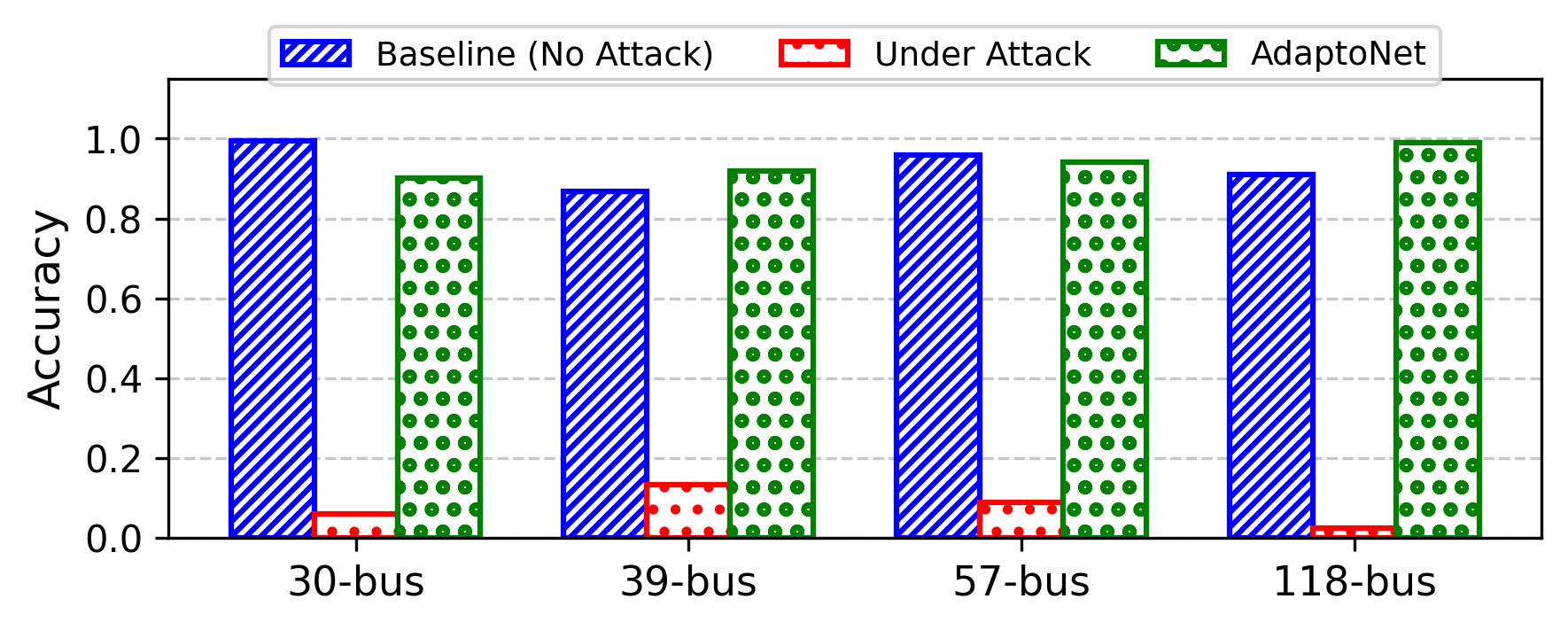}
    \captionof{figure}{Type~A accuracy comparison. AdaptoNet recovers accuracy from below 13\% (under attack) to 90--99\%, approaching the 87--99\% baseline.}
    \label{fig:typeA_acc}
  \end{minipage}
  \hfill
  \begin{minipage}{0.48\textwidth}
    \centering
    \includegraphics[width=\textwidth, height=0.35\textwidth]{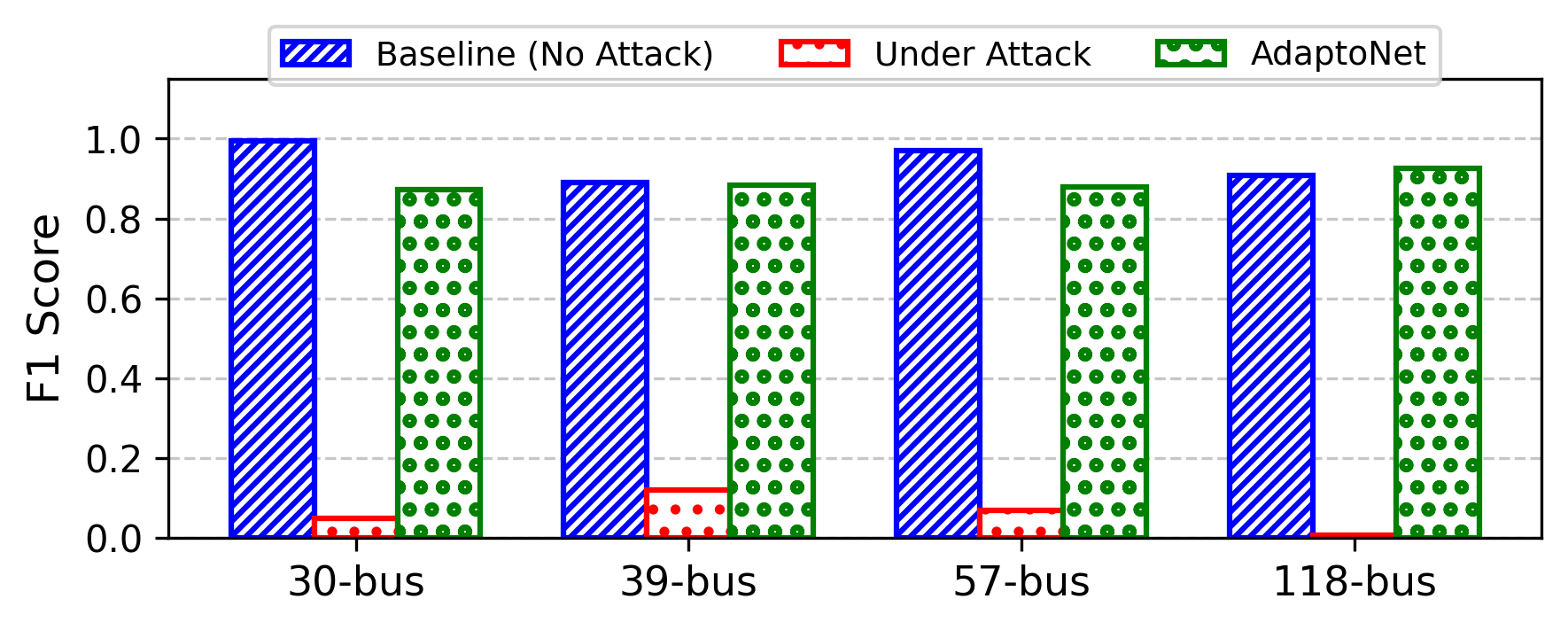}
    \captionof{figure}{Type~A F1-score comparison. AdaptoNet recovers F1-score from below 12\% (under attack) to 87--93\%, approaching the 89--99\% baseline.}
    \label{fig:typeA_f1}
  \end{minipage}
  \vspace{5pt}
  \begin{minipage}{0.48\textwidth}
    \centering
    \includegraphics[width=\textwidth, height=0.35\textwidth]{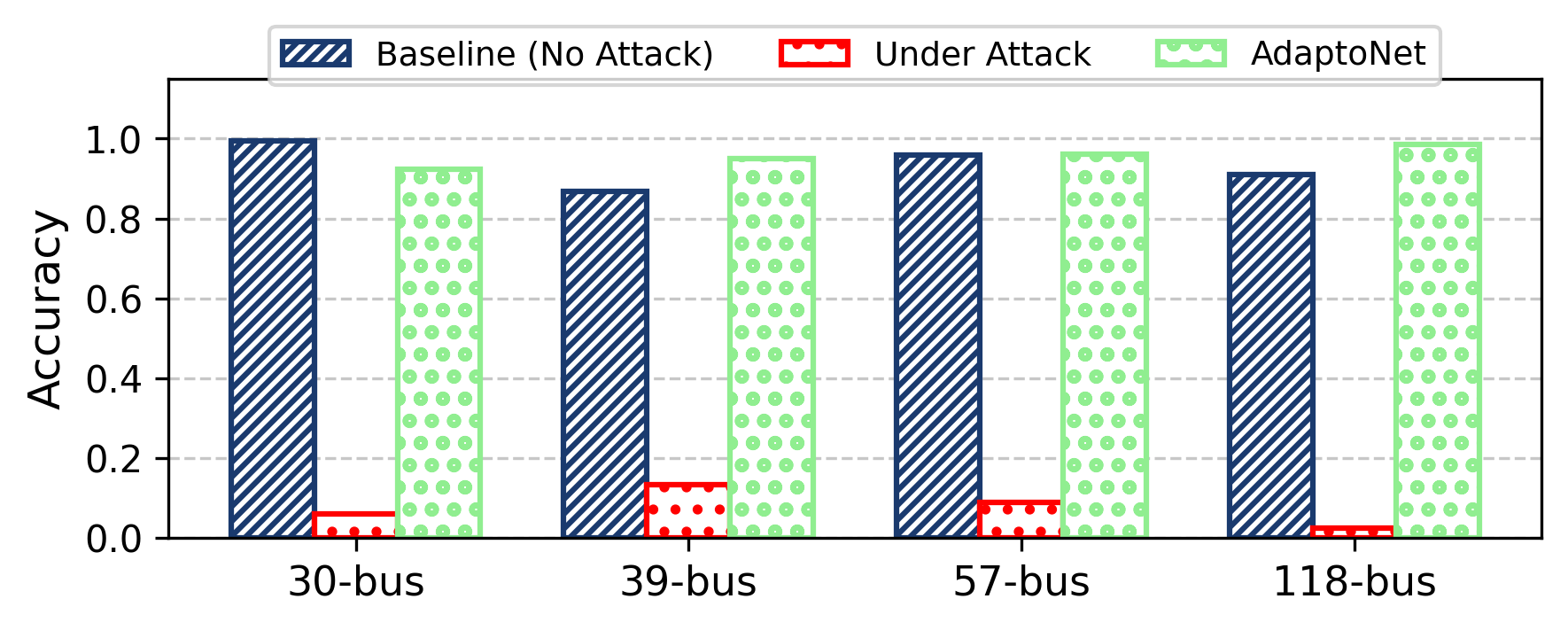}
    \captionof{figure}{Type~B accuracy comparison. Under larger attack regions (20\% compromised), AdaptoNet maintains 92--99\% accuracy.}
    \label{fig:typeB_acc}
  \end{minipage}
  \hfill
  \begin{minipage}{0.48\textwidth}
    \centering
    \includegraphics[width=\textwidth, height=0.35\textwidth]{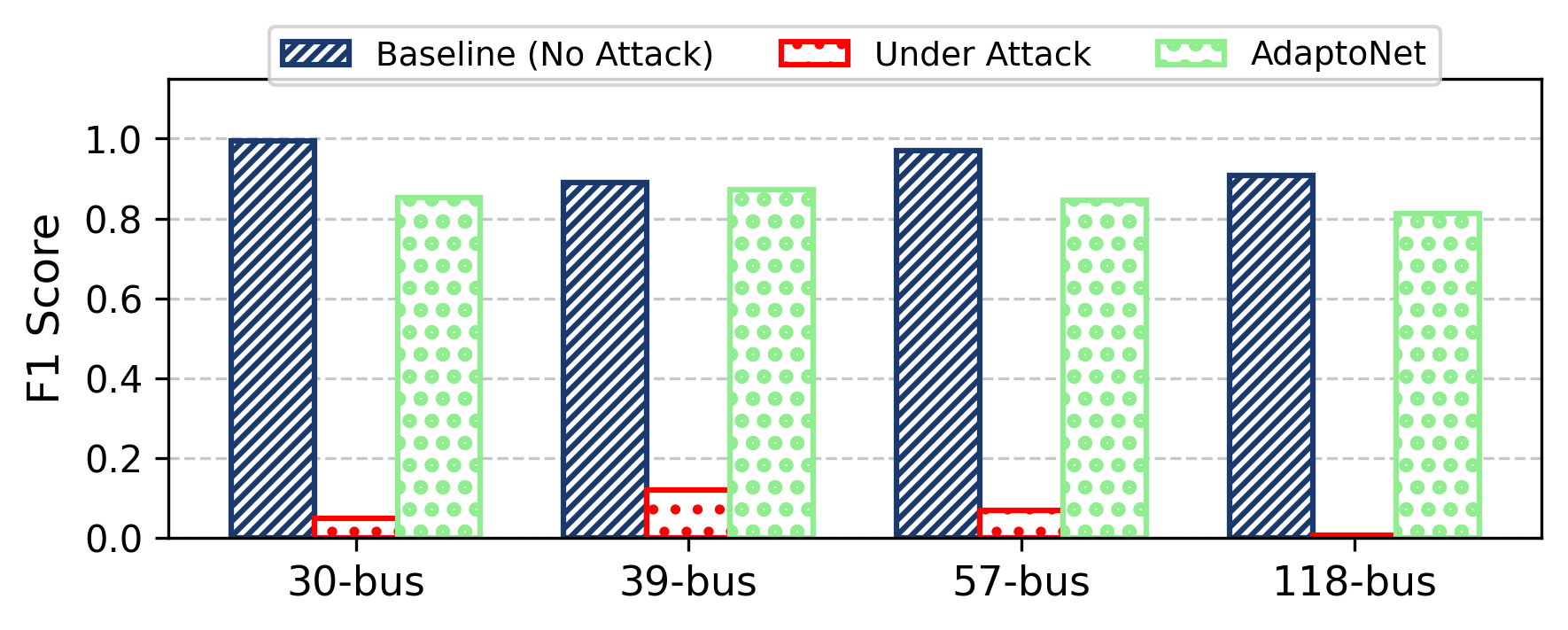}
    \captionof{figure}{Type~B F1-score comparison. With 20\% compromised buses, AdaptoNet achieves 81--87\% F1-score.}
    \label{fig:typeB_f1}
  \end{minipage}
  \vspace{-20pt}
\end{figure*}

Figs.~\ref{fig:typeA_acc}--\ref{fig:typeB_f1} compare F1 and accuracy across all four systems, six evaluation cases, and both region types.
AdaptoNet recovers F1 from below 12\% to 87\%--92.5\% and accuracy from below 13\% to 90\%--99\% across all four systems, an approximate sevenfold improvement, approaching the 89\%--99\% baseline achieved with complete measurements. Recall is consistently high (88\%--96.8\% for Type~A), indicating the model catches most outages, a desirable bias for security applications where missing a real outage is costlier than a false alarm. Under Type~B regions (20\% compromised), F1 ranges from 81.2\%--87.3\% and accuracy from 92.4\%--98.5\%, with precision dropping more than recall as larger denied regions increase false positives among topologically similar lines. Note that accuracy is high across all systems due to class imbalance: with single-line outages, most lines are correctly classified as behaving normally (true negatives), inflating accuracy. F1 is the primary metric, as it better captures detection performance for line outages.

\begin{table}[htbp]
  \centering
  \caption{AdaptoNet performance under in-region attack, averaged across all regions and six evaluation cases.}
  \label{tab:adaptonet_results}
  \setlength{\tabcolsep}{3pt}
  \begin{tabular}{C{0.5in} C{0.5in} C{0.4in} C{0.4in} C{0.4in} C{0.4in}}
  \toprule
  \textbf{Type} & \textbf{System} & \textbf{Acc.} & \textbf{F1} & \textbf{Prec.} & \textbf{Recall} \\
  \midrule
  \multirow{4}{*}{\textbf{A}} & 30-bus  & .900 & .872 & .839 & .943 \\
                     & 39-bus  & .918 & .884 & .918 & .882 \\
                     & 57-bus  & .941 & .878 & .875 & .909 \\
                     & 118-bus & .989 & .925 & .892 & .968 \\
  \midrule
  \multirow{4}{*}{\textbf{B}} & 30-bus  & .924 & .852 & .853 & .876 \\
      & 39-bus  & .951 & .873 & .901 & .859 \\
      & 57-bus  & .962 & .845 & .793 & .920 \\
      & 118-bus & .985 & .812 & .736 & .934 \\
  \bottomrule
  \end{tabular}
\end{table}

\subsubsection{Key Findings}

\textbf{Two-stage training preserves foundational knowledge.} The foundational module achieves 89\%--99\% F1 under complete measurement conditions. When frozen and paired with the adaptive module, AdaptoNet recovers to 81\%--93\% F1 under attack, compared to the baseline CNN's collapse to below 12\% while under attack. This demonstrates that the two-stage approach preserves learned grid physics while enabling attack-specific adaptation. 

\textbf{Conditioning drives awareness.} The binary availability mask embeds spatial indicators of measurement availability into the model, giving it the ability to capture the physical system's state and disruptions effectively. 



\textbf{Performance is competitive across system sizes.} Under Type~A attacks, larger systems tend to achieve higher F1 (0.925 on 118-bus vs.\ 0.872 on 30-bus), as more inside lines per region provide richer classification signals. This trend is not monotonic: under the larger Type~B denial, the 118-bus system yields the lowest F1 (0.812), so the relationship with system size is conditional rather than uniform. Per-region analysis shows most regions exceed 0.95 F1, with the lower mean driven by a small number of outlier regions of topologically adjacent lines that produce nearly identical fault signatures (e.g., the region spanning buses 26, 28, and 29 in the 39-bus system, which falls to roughly 0.52 F1). Excluding these few outlier regions raises the mean F1 by several points. When lines are this indistinguishable, an operator would treat them as a single protection group and act on the affected zone rather than on an individual line. The inclusive means in Table~\ref{tab:adaptonet_results} values include all regions; excluding the outliers only shows how much they pull the mean down.

\textbf{Detection latency supports real-time deployment.} Table~{\ref{tab:latency}} reports inference latency per forward pass and end-to-end prediction on GPU and CPU (1-thread and 4-thread). AdaptoNet completes a detection in 1.2--1.72 ms on NVIDIA TITAN RTX GPU across all four systems. On CPU, we test the most conservative setting (single-threaded) and a more realistic setting (4-threaded). Single-threaded inference ranges from 1.8 ms (30-bus) to 54.9 ms (118-bus), with 4-threaded execution reducing the 118-bus case to 18.5 ms. These results support deployment on both GPU-accelerated control centers and CPU-only edge devices within wide-area monitoring cycles. As a baseline comparison, all results are within the PMU reporting cycle of 10--60 Hz, which indicates the approach scales to larger grids without exceeding the cycle.
\begin{table}[htbp]
\centering
\caption{AdaptoNet inference latency per bus system (ms, mean over all four IEEE systems).}
\label{tab:latency}
\begin{tabular}{lcccccc}
\toprule
 & \multicolumn{2}{c}{GPU} & \multicolumn{2}{c}{CPU, 1 thr} & \multicolumn{2}{c}{CPU, 4 thr} \\
\cmidrule(lr){2-3} \cmidrule(lr){4-5} \cmidrule(lr){6-7}
System & Fwd & E2E & Fwd & E2E & Fwd & E2E \\
\midrule
30-bus  & 1.20 & 1.46 & 1.81 & 1.90 & 1.37 & 1.41 \\
39-bus  & 1.20 & 1.47 & 2.50 & 2.59 & 1.66 & 1.71 \\
57-bus  & 1.27 & 1.47 & 3.20 & 3.18 & 2.08 & 2.14 \\
118-bus & 1.46 & 1.72 & 54.90 & 54.68 & 18.46 & 18.55 \\
\bottomrule
\end{tabular}
\end{table}

\subsection{Discussion}\label{subsec:discuss}
This paper focuses on AdaptoNet's capability to restore detection performance under in-region cyber-physical attacks by using a modular framework and the data denial state as an observable input. The design principle can be applied to a wide range of attack scenarios and operational conditions. 
\textbf{Attack Scenario Extension.} In-region attacks follow existing attack scenario considered in~\cite{After_attack_Bayes, lineoutage_PMU_joint_attack}. 
Unlike those studies, AdaptoNet presents no such restrictions on these attacks. The mask variables, directly indicating data availability, can specify arbitrary attack scenarios, regardless of whether data is compromised outside the line-outage region (e.g., in the out-region or random attacks). Meanwhile, the output layer of AdaptoNet specifically indicates branch status, requiring no structural changes when we apply AdaptoNet in multi-line outage scenarios.  

\textbf{Missing Data Imputation Complement.} 
AdaptoNet is neither a missing data imputation method nor does it rely on them. It introduces structural novelty in neural networks to increase their resilience to missing data. However, AdaptoNet can be integrated with existing data imputation methods. Instead of using zeroed values in the missing data, we can use predicted ones and use a mask variable to explicitly distinguish predicted and real values during training.


\section{Conclusion}\label{conclusions}

We proposed AdaptoNet, a modular neural network that adapts to measurement availability for detecting line outages during cyber-physical attacks. By pairing a foundational module with a trainable adaptive module conditioned on measurement availability, AdaptoNet recovers F1 from below 12\% to above 81\% and accuracy from below 13\% to 90\%--99\% across four IEEE test systems, approaching the 89\%--99\% baseline with complete measurements. Conditioning embeds spatial information on measurement availability, the modular split preserves learned grid knowledge during adaptation, and performance remains competitive across the four system sizes.

In future work, we plan to expand AdaptoNet to more complicated attack scenarios, including different data compromises and multi-line outages, investigate AdaptoNet's generation to different neural network infrastructures, and study its potential integration with other data imputation methods.

\section*{Acknowledgment}

This material is based upon work partially supported by the National Science Foundation (NSF) under Grant CNS-2144513 and CNS-2247722, the
Department of Energy (DOE) under Award DE-CR0000039,  and the Office of Naval Research (ONR) under Contract no. N000142412129. Any opinions, findings, and conclusions or recommendations expressed in this material are those of the author(s) and do not necessarily reflect the view of the NSF, the DOE, and the ONR.

\bibliographystyle{IEEEtran}
\bibliography{SGC_references}

@techreport{lee2016analysis,
  title     = {Analysis of the cyber attack on the {Ukrainian} power grid},
  institution = {{SANS} and E-{ISAC}},
  author    = {Lee, Robert M. and Assante, Michael J. and Conway, Tim},
  year      = {2016}
}

@article{proska2023sandworm,
  title   = {Sandworm Disrupts Power in {Ukraine} Using a Novel Attack Against Operational Technology},
  author  = {Proska, Ken and Wolfram, John and Wilson, Jared and Black, Dan and Lunden, Keith
             and Zafra, Daniel Kapellmann and Brubaker, Nathan and Mclellan, Tyler and Sistrunk, Chris},
  journal = {Mandiant},
  year    = {2023}
}

@article{stekhoven2011missforest,
  author = {Stekhoven, Daniel J. and Bühlmann, Peter},
    title = {MissForest—non-parametric missing value imputation for mixed-type data},
    journal = {Bioinformatics},
    volume = {28},
    number = {1},
    pages = {112-118},
    year = {2012},
    month = {01},
    issn = {1367-4803},
    doi = {10.1093/bioinformatics/btr597}
}

@misc{collier2021vae,
  title         = {{VAEs} in the Presence of Missing Data},
  author        = {Mark Collier and Alfredo Nazabal and Christopher K. I. Williams},
  year          = {2021},
  eprint        = {2006.05301},
  archivePrefix = {arXiv},
  primaryClass  = {cs.LG},
  url           = {https://arxiv.org/abs/2006.05301}
}

@InProceedings{yoon2018gain,
  title     = {{GAIN}: Missing Data Imputation using Generative Adversarial Nets},
  author    = {Yoon, Jinsung and Jordon, James and van der Schaar, Mihaela},
  booktitle = {Proceedings of the 35th International Conference on Machine Learning},
  year      = {2018},
  volume    = {80},
  month     = {Jul}
}

@INPROCEEDINGS{xu2025pseudo,
  author={Xu, Tao and Wang, Kaiqi and Zhang, Jiadong and Qiao, Ji and Zhao, Zixuan and Zhu, Hong and Sun, Kai},
  booktitle={2025 IEEE Power \& Energy Society General Meeting (PESGM)}, 
  title={Pseudo-Measurement Enhancement in Power Distribution Systems}, 
  year={2025},
  volume={},
  number={},
  doi={10.1109/PESGM52009.2025.11225469}
  }

@article{Learning_to_infer,
  author={Zhao, Yue and Chen, Jianshu and Poor, H. Vincent},
  journal={IEEE Transactions on Smart Grid}, 
  title={A Learning-to-Infer Method for Real-Time Power Grid Multi-Line Outage Identification}, 
  year={2020},
  volume={11},
  number={1},
  pages={555-564},
  doi={10.1109/TSG.2019.2925405}
}

@ARTICLE{Lineoutages_PMU,
  author  = {Tate, Joseph Euzebe and Overbye, Thomas J.},
  journal = {IEEE Transactions on Power Systems},
  title   = {Line Outage Detection Using Phasor Angle Measurements},
  year    = {2008},
  volume  = {23},
  number  = {4},
  pages   = {1644--1652},
  doi     = {10.1109/TPWRS.2008.2004826}
}

@article{CNN_LSTM,
  title   = {Hybrid {CNN-LSTM} approaches for identification of type and locations
             of transmission line faults},
  journal = {International Journal of Electrical Power \& Energy Systems},
  volume  = {135},
  pages   = {107563},
  year    = {2022},
  doi     = {https://doi.org/10.1016/j.ijepes.2021.107563},
  author  = {Arash Moradzadeh and Hamid Teimourzadeh and Behnam Mohammadi-Ivatloo
             and Kazem Pourhossein}
}

@INPROCEEDINGS{CNN_faultdiagnosis,
  author    = {Hassani, Hossein and Farajzadeh-Zanjani, Maryam and Razavi-Far, Roozbeh
               and Saif, Mehrdad and Palade, Vasile},
  booktitle = {2019 18th IEEE International Conference On Machine Learning And Applications (ICMLA)},
  title     = {Design of a Cost-Effective Deep Convolutional Neural Network--Based Scheme
               for Diagnosing Faults in Smart Grids},
  year      = {2019},
  pages     = {1420--1425},
  doi       = {10.1109/ICMLA.2019.00232}
}

@INPROCEEDINGS{previous_work,
  author    = {Rogers, Jennifer and Danilczyk, William and Lin, Hui and Sun, Yan Lindsay},
  booktitle = {2023 IEEE 11th International Conference on Smart Energy Grid Engineering (SEGE)},
  title     = {Learning from Future: Prediction-based Data Augmentation to Enhance
               Power Grids Fault Detection},
  year      = {2023}
}

@ARTICLE{Hidden_markov,
  author  = {Huang, Qingqing and Shao, Leilai and Li, Na},
  journal = {IEEE Transactions on Power Systems},
  title   = {Dynamic Detection of Transmission Line Outages Using Hidden Markov Models},
  year    = {2016},
  volume  = {31},
  number  = {3},
  pages   = {2026--2033},
  doi     = {10.1109/TPWRS.2015.2456852}
}

@ARTICLE{sysID,
  author  = {Babakmehr, Mohammad and Harirchi, Farnaz and Al-Durra, Ahmed and
             Muyeen, S. M. and Sim{\~o}es, Marcelo Godoy},
  journal = {IEEE Transactions on Industry Applications},
  title   = {Compressive System Identification for Multiple Line Outage Detection
             in Smart Grids},
  year    = {2019},
  volume  = {55},
  number  = {5},
  pages   = {4462--4473},
  doi     = {10.1109/TIA.2019.2921260}
}

@article{Zhuo_CNN_2024,
  title     = {Distribution grid fault classification and localization using
               convolutional neural networks},
  author    = {Zhou, Ming and Kazemi, Nazli and Musilek, Petr},
  journal   = {Smart Grids and Sustainable Energy},
  volume    = {9},
  number    = {1},
  pages     = {24},
  year      = {2024},
  publisher = {Springer}
}

@misc{shakiba2022transferlearningfaultdiagnosis,
  title         = {Transfer Learning for Fault Diagnosis of Transmission Lines},
  author        = {Fatemeh Mohammadi Shakiba and Milad Shojaee and S. Mohsen Azizi
                   and Mengchu Zhou},
  year          = {2022},
  eprint        = {2201.08018},
  archivePrefix = {arXiv},
  primaryClass  = {cs.LG},
  url           = {https://arxiv.org/abs/2201.08018}
}

@inproceedings{rogers2024prediction,
  title        = {Prediction-Based Data Augmentation for Smart Grid Line Outage Detection},
  author       = {Rogers, Jennifer and Lin, Hui and Sun, Yan Lindsay},
  booktitle    = {2024 56th North American Power Symposium (NAPS)},
  pages        = {1--6},
  year         = {2024},
  organization = {IEEE}
}

@article{cyber_attack_types,
  title   = {Detection of power grid disturbances and cyber-attacks based on machine learning},
  journal = {Journal of Information Security and Applications},
  volume  = {46},
  pages   = {42--52},
  year    = {2019},
  doi     = {https://doi.org/10.1016/j.jisa.2019.02.008},
  author  = {Defu Wang and Xiaojuan Wang and Yong Zhang and Lei Jin}
}

@article{coordinated_attack,
  title   = {A tri-level optimization model to mitigate coordinated attacks on electric
             power systems in a cyber-physical environment},
  journal = {Applied Energy},
  volume  = {235},
  year    = {2019},
  doi     = {https://doi.org/10.1016/j.apenergy.2018.10.077},
  author  = {Kexing Lai and Mahesh Illindala and Karthikeyan Subramaniam}
}

@ARTICLE{After_attack_Bayes,
  author  = {Soltan, Saleh and Mittal, Prateek and Poor, H. Vincent},
  journal = {IEEE Transactions on Power Systems},
  title   = {Line Failure Detection After a Cyber-Physical Attack on the Grid
             Using Bayesian Regression},
  year    = {2019},
  volume  = {34},
  number  = {5},
  pages   = {3758--3768},
  doi     = {10.1109/TPWRS.2019.2910396}
}

@INPROCEEDINGS{lineoutage_PMU_joint_attack,
  author    = {Hossain, MD Jakir and Rahnamay-Naeini, Mahshid},
  booktitle = {2019 IEEE Power \& Energy Society General Meeting (PESGM)},
  title     = {Line Failure Detection from PMU Data after a Joint Cyber-Physical Attack},
  year      = {2019},
  pages     = {1--5},
  doi       = {10.1109/PESGM40551.2019.8973656}
}

@ARTICLE{soltan2017react,
  author={Soltan, Saleh and Yannakakis, Mihalis and Zussman, Gil},
  journal={IEEE Transactions on Network Science and Engineering}, 
  title={REACT to Cyber Attacks on Power Grids}, 
  year={2019},
  volume={6},
  number={3},
  pages={459--473},
  doi={10.1109/TNSE.2018.2837894}
}

@misc{ren2025faultlocalizationstateestimation,
  title         = {Fault Localization and State Estimation of Power Grid under
                   Parallel Cyber-Physical Attacks},
  author        = {Junhao Ren and Kai Zhao and Guangxiao Zhang and Xinghua Liu
                   and Chao Zhai and Gaoxi Xiao},
  year          = {2025},
  eprint        = {2503.05797},
  archivePrefix = {arXiv},
  primaryClass  = {eess.SY},
  url = {https://arxiv.org/abs/2503.05797},
}

@article{zhang2021smart,
  author  = {Zhang, Hang and Liu, Bo and Wu, Hongyu},
  journal = {IEEE Access},
  title   = {Smart Grid Cyber-Physical Attack and Defense: A Review},
  year    = {2021},
  volume  = {9},
  pages   = {29641--29659},
  doi     = {10.1109/ACCESS.2021.3058628}
}

@article{liu2011fdia,
  author  = {Liu, Yao and Ning, Peng and Reiter, Michael K.},
  title   = {False Data Injection Attacks against State Estimation in Electric Power Grids},
  journal = {ACM Transactions on Information and System Security (TISSEC)},
  volume  = {14},
  number  = {1},
  pages   = {1--33},
  year    = {2011},
  doi     = {10.1145/1952982.1952995}
}

@ARTICLE{kosut2011malicious,
  author  = {Kosut, Oliver and Jia, Liyan and Thomas, Robert J. and Tong, Lang},
  journal = {IEEE Transactions on Smart Grid},
  title   = {Malicious Data Attacks on the Smart Grid},
  year    = {2011},
  volume  = {2},
  number  = {4},
  pages   = {645--658},
  doi     = {10.1109/TSG.2011.2163807}
}

@ARTICLE{lin2018runtime,
  author  = {Lin, Hui and Slagell, Adam and Kalbarczyk, Zbigniew T. and Sauer, Peter W.
             and Iyer, Ravishankar K.},
  journal = {IEEE Transactions on Smart Grid},
  title   = {Runtime Semantic Security Analysis to Detect and Mitigate Control-Related
             Attacks in Power Grids},
  year    = {2018},
  volume  = {9},
  number  = {1},
  pages   = {163--178},
  doi     = {10.1109/TSG.2016.2547742}
}

@inproceedings{lin2016safetycritical,
  location  = {New York, NY, USA},
  title     = {Safety-critical Cyber-physical Attacks: Analysis, Detection, and Mitigation},
  doi       = {10.1145/2898375.2898391},
  booktitle = {Proceedings of the Symposium and Bootcamp on the Science of Security (HotSoS)},
  author    = {Lin, Hui and Alemzadeh, Homa and Chen, Daniel and Kalbarczyk, Zbigniew
               and Iyer, Ravishankar K.},
  year      = {2016}
}

@InProceedings{rezazad2019detecting,
  author    = {Rezazad, Mostafa and Brust, Matthias R. and Akbari, Mohammad
               and Bouvry, Pascal and Cheung, Ngai-Man},
  title     = {Detecting Target-Area Link-Flooding DDoS Attacks Using Traffic
               Analysis and Supervised Learning},
  booktitle = {Advances in Information and Communication Networks},
  year      = {2019},
  publisher = {Springer International Publishing},
  pages     = {180--202}
}

@inproceedings{zhang2023adding,
  author    = {Zhang, Lvmin and Rao, Anyi and Agrawala, Maneesh},
  booktitle = {2023 IEEE/CVF International Conference on Computer Vision (ICCV)},
  title     = {Adding Conditional Control to Text-to-Image Diffusion Models},
  year      = {2023},
  pages     = {3813--3824},
  doi       = {10.1109/ICCV51070.2023.00355}
}

@inproceedings{Conditioning_image_generation,
  author    = {Amir Hertz and Ron Mokady and Jay Tenenbaum and Kfir Aberman
               and Yael Pritch and Daniel Cohen-Or},
  title     = {Prompt-to-Prompt Image Editing with Cross Attention Control},
  booktitle = {Proceedings of the International Conference on Learning Representations (ICLR)},
  year      = {2023}
}

@inproceedings{nn_conditioning,
  title        = {Conditioning neural networks: A case study of electricity load forecasting},
  author       = {Hosseini, Hossein and Hooshmand, Ali and Sharma, Ratnesh},
  booktitle    = {2018 IEEE International Conference on Big Data (Big Data)},
  year         = {2018},
}

@ARTICLE{matpower,
  author  = {Zimmerman, Ray Daniel and Murillo-S{\'a}nchez, Carlos Edmundo
             and Thomas, Robert John},
  journal = {IEEE Transactions on Power Systems},
  title   = {{MATPOWER}: Steady-State Operations, Planning, and Analysis Tools
             for Power Systems Research and Education},
  year    = {2011},
  volume  = {26},
  number  = {1},
  pages   = {12--19},
  doi     = {10.1109/TPWRS.2010.2051168}
}

@ARTICLE{b20,
  author  = {Birchfield, Adam B. and Xu, Ti and Gegner, Kathleen M.
             and Shetye, Komal S. and Overbye, Thomas J.},
  journal = {IEEE Transactions on Power Systems},
  title   = {Grid Structural Characteristics as Validation Criteria for Synthetic Networks},
  year    = {2017},
  volume  = {32},
  number  = {4},
  pages   = {3258--3265},
  doi     = {10.1109/TPWRS.2016.2616385}
}

@article{mcroberts2018improving,
  author = {McRoberts, D. Brent and Quiring, Steven M. and Guikema, Seth D.},
title = {Improving Hurricane Power Outage Prediction Models Through the Inclusion of Local Environmental Factors},
journal = {Risk Analysis},
volume = {38},
number = {12},
doi = {https://doi.org/10.1111/risa.12728},
year = {2018}
}

\end{document}